\documentclass[letterpaper, 10 pt, conference]{ieeeconf}

\IEEEoverridecommandlockouts                              
\usepackage{amsmath}
\usepackage{mathtools}
\usepackage{amssymb}
\usepackage{subcaption}
\usepackage{xcolor}
\usepackage{graphicx}
\usepackage{optidef}
\usepackage{enumerate}
\usepackage{multirow}
\usepackage[linesnumbered,ruled,vlined]{algorithm2e}
\usepackage{algorithmicx}
\usepackage[utf8]{inputenc}
\usepackage{booktabs} 
\usepackage{tabularx} 

\newtheorem{assumption}{\textbf{Assumption}}[section]
\newtheorem{problem}{\textbf{Problem}}[section]
\newtheorem{remark}{\textbf{Remark}}[section]
\newtheorem{lemma}{\textbf{Lemma}}[section]
\newtheorem{theorem}{\textbf{Theorem}}[section]

\newtheorem{definition}{\textbf{Definition}}[section]

\usepackage{balance}

\title{\LARGE \bf
Resilient Learning-Based Control Under Denial-of-Service Attacks
}

\author{Sayan Chakraborty$^{1}$, Weinan Gao$^{2}$, Kyriakos G. Vamvoudakis$^{3}$, Zhong-Ping Jiang$^{1}$
\thanks{$^{1}$S. Chakraborty and Z.-P. Jiang are with the CAN Lab, New York University, Brooklyn, NY 11201 USA, {\tt\small sc8804@nyu.edu, zjiang@nyu.edu}}%
\thanks{$^{2}$W. Gao is with the State Key Laboratory of Synthetical Automation for Process Industries, Northeastern University, Shenyang 110819, China, {\tt\small gaown@mail.neu.edu.cn}}%
\thanks{$^{3}$K. G. Vamvoudakis is with The Daniel Guggenheim School of Aerospace Engineering, Georgia Institute of Technology, GA 30332-0150 USA,  {\tt\small kyriakos@gatech.edu}}%
\thanks{This work was supported in part by the NSF under grant nos. CNS-$2148309$, EPCN-$2210320$, CPS-$2227185$, S\&AS-$1849198$,  CPS-$1851588$, and CPS-$2227153$, and by Minerva under grant No. N$00014-18-1-2874$.}
}

\begin{document}

\maketitle
\thispagestyle{empty}
\pagestyle{empty}

\begin{abstract}
In this paper, we have proposed a resilient reinforcement learning method for discrete-time linear systems with unknown parameters, under denial-of-service (DoS) attacks. The proposed method is based on policy iteration that learns the optimal controller from input-state data amidst DoS attacks. We achieve an  upper bound for the DoS duration to ensure closed-loop stability. The resilience of the closed-loop system, when subjected to DoS attacks with the learned controller and an internal model, has been thoroughly examined. The effectiveness of the proposed methodology is demonstrated on an inverted pendulum on a cart.
\end{abstract}

\section{INTRODUCTION}

Reinforcement learning (RL) outlines strategies for an agent to adjust its actions when interacting with an unfamiliar environment, aiming to fulfill a long-term objective \cite{sutton2018reinforcement}. Researchers from the control community have used ideas from RL and adaptive/approximate dynamic programming (ADP) \cite{bertsekas2012dynamic,bellman1966dynamic, lewis2012reinforcement, werbos1974beyond} to develop data-driven adaptive optimal control methods to address the stabilization problem of dynamical systems (see \cite{jiang2012computational, vrabie2009adaptive, vamvoudakis2020synchronous, chakraborty2022automated, jiang2014adaptive, jiang2014robust, lewis2009reinforcement}). As a generalization, the authors in \cite{gao2016adaptive} combined ADP with output regulation theory for asymptotic tracking and disturbance rejection, later subsequently extending this framework to a data-driven approach for non-linear systems in \cite{gao2017learningg}. This approach has been extended to multi-agent systems in \cite{gao2021reinforcement, odekunle2020reinforcement} and references therein.

However, the existing ADP studies usually rely on the assumption that communication channels for control and measurement are ideal, which makes the designed controller vulnerable to cyberattacks. Consequently, it becomes crucial to extend the analysis of control systems beyond stability and robustness to include resilience, ensuring that systems can effectively withstand and recover from cyber threats. When a system is under DoS attack, the transmission of information is blocked among networks \cite{amin2009safe, teixeira2015secure}. An explicit characterization of DoS frequency and duration has been given in \cite{de2015input} such that the closed-loop system remains robustly stable under DoS attacks. Other research in this direction can be found in \cite{deng2020distributed, an2018decentralized, zhang2019resilient, zhang2023secure}. However, the authors in the aforementioned references consider neither model-free controller design nor resilient analysis for the closed-loop system under DoS attacks. Most recently, learning-based approaches have been adopted in \cite{gao2022resilient, galarza2021extremum, zhai2021data, shi2023distributed} by using RL, ADP, and extremum seeking to defend the closed-loop system under adversarial attacks. 

Most of the above-mentioned works have considered continuous-time systems. Recently, researchers have begun to explore learning-based approaches to study the resilience of closed-loop discrete-time systems under DoS attacks. In order to guarantee resilience under DoS attacks, many authors have adopted data-driven predictive control \cite{liu2022data, zhang2022robust} and adaptive control \cite{ma2021distributed, li2022learning} techniques to address the problem of learning-based control to guarantee the resilience of closed-loop discrete-time systems under DoS attacks. In this paper, we address the data-driven optimal output regulation problem for discrete-time systems with unknown parameters under DoS attacks using RL and ADP. The problem of output regulation focuses on the development of a feedback control law that ensures asymptotic tracking and disturbance rejection. To our knowledge, learning-based optimal output regulation for discrete-time systems under DoS attacks has been proposed for the first time. Our approach enables direct analysis of the closed-loop system’s resilience with the optimal controller and achieves an upper bound for the DoS attack duration that the system can withstand while maintaining stability

The rest of the paper is organized as follows. Section~\ref{sec:prelim} formulates the control objective and controller design in the absence of DoS attacks. Section~\ref{sec:resAnalysis} provides the resilience analysis of the closed-loop system under DoS attacks. Section~\ref{sec:learn} provides the online learning method using policy iteration when the system is under DoS attacks. Section~\ref{sec:sim} presents the simulation results. Finally, the conclusion and future work are mentioned in Section~\ref{sec:conc}.

\textbf{Facts and notations:} $\mathbb{R}_{+}$ denotes the set of non-negative real numbers. $\mathbb{Z}_{+}$ the set of non-negative integers. $\lvert x\rvert$ denotes the Euclidean norm of a vector $x\in\mathbb{R}^{n}$. $\lvert A\rvert$ denotes the induced matrix norm for a matrix $A \in \mathbb{R}^{m\times n}$. For a square matrix $A$, $\sigma(A)$ denotes the spectrum of $A$. For a real symmetric matrix $A$, $\lambda_{m}(A)$ and $\lambda_{M}(A)$ denote the minimum and maximum eigenvalues of $A$, respectively. For a symmetric positive definite matrix $P \in \mathbb{R}^{m \times m}$ and $x \in \mathbb{R}^{m}$, we have $ \lambda_{m}(P)\lvert x \vert^{2} \leq x^\mathrm{T}Px \leq \lambda_{M}(P)\lvert x \vert^{2}$. For any function $\zeta\colon \mathbb{Z}_{+}\rightarrow \mathbb{R}^{n}$, $\lVert\zeta\rVert = \text{sup}\{ \lvert \zeta_{k} \rvert\colon k\in\mathbb{Z}_{+} \}\leq \infty$. A function $\alpha\colon\mathbb{R}_{+}\rightarrow\mathbb{R}_{+}$ belongs to class $\mathcal{K}$ if it is continuous, strictly increasing and $\alpha(0)=0$. A function $\alpha$ belongs to class $\mathcal{K}_{\infty}$ if it is of class $\mathcal{K}$ and also $\alpha(r)\rightarrow \infty$ as $r\rightarrow\infty$. A function $\beta\colon\mathbb{R}_{+}\times\mathbb{R}_{+}\rightarrow\mathbb{R}_{+}$ belongs to class $\mathcal{KL}$ if, for each fixed $k\geq0$, the function $\beta(.,k)$ belongs to class $\mathcal{K}$, and for each fixed $r\geq0$, the function $\beta(r,.)$ is decreasing and $\beta(r,k)\rightarrow0$ as $k\rightarrow\infty$. For any $x$, $y$ $\in \mathbb{R}^{n}$, and for any $\epsilon > 0$, we have $x^\mathrm{T}y \leq \frac{1}{4\epsilon}x^\mathrm{T}x + \epsilon y^\mathrm{T}y$. $\otimes$ indicates the Kronecker product, $\text{vec}(T) = \begin{bmatrix} t_{1}^\mathrm{T}, t_{2}^\mathrm{T}, \cdots, t_{m}^\mathrm{T} \end{bmatrix}^\mathrm{T}$ with $t_{i} \in \mathbb{R}^{r}$ being the columns of $T \in \mathbb{R}^{r\times m}$. For a symmetric matrix $P \in \mathbb{R}^{m \times m}$, $\text{vecs}(P) = [ p_{11}, 2p_{12},\cdots,2p_{1m}, p_{22}, 2p_{23}, \cdots, 2p_{(m-1)m}, p_{mm} ]^\mathrm{T} \in \mathbb{R}^{(1/2)m(m+1)}$, for a column vector $v \in \mathbb{R}^{n}$, $ \text{vecv}(v) = [ v_{1}^{2}, v_{1}v_{2}, \cdots, v_{1}v_{n}, v_{2}^{2}, v_{2}v_{3}, \cdots, v_{n-1}v_{n}, v_{n}^{2} ]^\mathrm{T} \in \mathbb{R}^{(1/2)n(n+1)}$.  For any two sequence of vectors $\{a_{i}\}_{i=k_{0}}^{k_{s}}$, $\{b_{i}\}_{i=k_{0}}^{k_{s}}$, define $\Xi_{a} = \big[\text{vecv}(a_{k_{0}+1})-\text{vecv}(a_{k_{0}}), \cdots, \text{vecv}(a_{k_{s}})-\text{vecv}(a_{k_{s}-1}) \big]^\mathrm{T}$, $J_{a,b} = \big[ a_{k_{0}}\otimes b_{k_{0}}, \cdots, a_{k_{s}}\otimes b_{k_{s}}\big]^\mathrm{T}$, $J_{a} = \big[ \text{vecv}(a_{k_{0}}), \cdots, \text{vecv}(a_{k_{s}})\big]^\mathrm{T}$. $I_{n}$ and $0_{n}$ is the identity matrix and the zero matrix of dimension $n \times n$, respectively.
\section{Preliminaries and problem formulation}\label{sec:prelim}

Consider the following discrete-time cascade system:
\begin{align}
x_{k+1} &= Ax_{k}+Bu_{k}+Dw_{k},  \label{eq:state}\\
w_{k+1} &= Ew_{k}, \label{eq:exoSys}\\
e_{k} &= Cx_{k} + Fw_{k}, \label{eq:error}
\end{align}
where $k \in \mathbb{Z}_{+}$, $A\in\mathbb{R}^{n\times n}$, $B\in\mathbb{R}^{n}$, $C \in \mathbb{R}^{1\times n}$, $D\in\mathbb{R}^{n\times q}$, $E \in \mathbb{R}^{q\times q}$, $F \in \mathbb{R}^{1\times q}$ are constant matrices, $e_{k}\in\mathbb{R}$ is the measurement output, $u_{k}\in\mathbb{R}$ the control input, $x_{k}\in\mathbb{R}^{n}$ is the state, $w_{k}\in\mathbb{R}^{q}$ is the state of the exosystem and  $y_{dk} = -F w_{k}$ is the reference signal.

\begin{assumption}\label{assum:1}
    The pair $(A,B)$ is stabilizable and the eigenvalues of $E$ are simple on the unit circle. \hfill  $\square$
\end{assumption}

\begin{assumption}\label{assum:2}
   $\mathrm{rank}\Bigg(\begin{bmatrix}
    A-\lambda I & B\\
    C & 0\\
    \end{bmatrix}\Bigg) = n + 1$, $\forall \lambda \in \sigma(E)$.  \hfill  $\square$
\end{assumption}

\begin{definition}\label{def:IntModel}
A dynamic compensator of the form
\begin{equation}
    z_{k+1} = \mathcal{G}_{1}z_{k} + \mathcal{G}_{2}e_{k},\quad \forall k \in \mathbb{Z}_{+} \label{eq:intModel}
\end{equation}
is called an internal model of the system \eqref{eq:state}-\eqref{eq:error} if the pair ($\mathcal{G}_{1}, \mathcal{G}_{2}$) incorporates an internal model of the exosystem matrix $E$ \cite{huang2004nonlinear}.\hfill  $\square$
\end{definition}

\begin{remark}
    In this work, we let $\mathcal{G}_{1} = E$, and choose $\mathcal{G}_{2}\in\mathbb{R}^{q}$ such that the pair ($E,\mathcal{G}_{2}$) is controllable.\hfill  $\square$
\end{remark}

In the absence of DoS attacks, we show in Lemma~\ref{lemma:1} that it is possible to develop a state-feedback controller for the discrete-time system \eqref{eq:state}-\eqref{eq:error} with an internal model \eqref{eq:intModel} that solves the output regulation problem. Consider the following augmented system $\forall k \in \mathbb{Z}_{+}$
\begin{align}
    x_{k+1} &= Ax_{k}+Bu_{k}+Dw_{k}, \nonumber\\
    w_{k+1} &= Ew_{k}, \nonumber \\
    e_{k} &= Cx_{k} + Fw_{k}, \nonumber\\
    z_{k+1} &= Ez_{k} + \mathcal{G}_{2}e_{k}. \label{eq:augSys1}
\end{align}

\begin{lemma}\label{lemma:1}
    Under Assumptions \ref{assum:1}-\ref{assum:2}, if there exists a state-feedback controller
    \begin{equation}
        u_{k} = -K_{x}x_{k} - K_{z}z_{k},\quad\forall k \in \mathbb{Z}_{+} \label{eq:feedContr}
    \end{equation}
    such that the closed-loop system matrix
    \begin{equation}
         A_{c} = \left[
            \begin{array}{cc}
            A-BK_{x} & -BK_{z}\\
            \mathcal{G}_{2}C & E\\
            \end{array}
            \right]
    \end{equation}
of the augmented system \eqref{eq:augSys1} is Schur. Then, the controller \eqref{eq:feedContr} solves the output regulation problem.
\end{lemma}

\begin{proof}
    Assumptions \ref{assum:1}-\ref{assum:2} guarantee the existence of a unique pair ($X,U$) solving the following regulator equations
    \begin{align}
        XE &= AX + BU +D, \label{eq:reg1}\\
        0 &= CX + F. \label{eq:reg2}
    \end{align}
    Also, \eqref{eq:reg2} combined with 
    \begin{align}
        XE &= (A-BK_{x})X - BK_{z}Z +D, \label{eq:reg3}\\
        ZE &= EZ + \mathcal{G}_{2}(CX + F), \label{eq:reg4}
    \end{align}
    have unique solutions $\hat{X}$ and $Z$ (see Lemma 1.38 in \cite{huang2004nonlinear}). This implies that  $X = \hat{X}$ and $U=-K_{x}X-K_{z}Z$. By defining the error states as $\tilde{x}_{k} = x_{k} - Xw_{k}$ and $\tilde{z}_{k} = z_{k} - Zw_{k}$, the following error system of \eqref{eq:augSys1} can be derived using \eqref{eq:reg3}-\eqref{eq:reg4}
    \begin{align}
        \tilde{x}_{k+1} &= (A-BK_{x}) \tilde{x}_{k} - BK_{z}\tilde{z}_{k}, \label{eq:errState1}\\
        \tilde{z}_{k+1} &= \mathcal{G}_{2}C\tilde{x}_{k} + E\tilde{z}_{k}. \label{eq:errState2}
    \end{align}
    Next, by defining $\tilde{\zeta}_{k} = [\tilde{x}^\mathrm{T}_{k}\quad\tilde{z}^\mathrm{T}_{k}]^\mathrm{T}\in\mathbb{R}^{n+q}$, and using \eqref{eq:errState1}-\eqref{eq:errState2} one can obtain
    \begin{align}
        \tilde{\zeta}_{k+1} &= A_{c}\tilde{\zeta}_{k},\\
        e_{k} &= \bar{C} \tilde{\zeta_{k}},
    \end{align}
    where $A_{c} = \bar{A}-\bar{B}K$, $\bar{C} = [C\quad0]$, $\bar{B} = [B^\mathrm{T}\quad0]^\mathrm{T}$, 
    and $\bar{A}=\left[
                        \begin{array}{cc}
                         A & 0\\
                        \mathcal{G}_{2}C & E
                        \end{array}
                        \right]$, $K = [K_{x}\quad K_{z}]$.
    
    Since $A_{c}$ is Schur, we have $\lim_{k\rightarrow\infty}\tilde{\zeta}_{k}=0$ and $\lim_{k\rightarrow\infty}e_{k}=0$. Thus, Lemma~\ref{lemma:1} is proved.                 
\end{proof}
As evident from Lemma~\ref{lemma:1}, the output regulation properties of \eqref{eq:augSys1} are guaranteed under a state-feedback controller of the form \eqref{eq:feedContr}. Furthermore, the optimal output regulation problem can be posed as follows.
\begin{problem}\label{prb:1}
    In order to obtain the optimal state-feedback controller that solves the output regulation problem for \eqref{eq:augSys1}, the following dynamic programming problem is solved
    \begin{align}
          \min_{\tilde{u}} \sum_{k=0}^{\infty} (\tilde{\zeta}_{k}^{\mathrm{T}}Q\tilde{\zeta}_{k}+\tilde{u}_{k}^{2})\\
        \text{s.t}\quad\tilde{\zeta}_{k+1} = \bar{A}\tilde{\zeta}_{k} + \bar{B}\tilde{u}_{k},
    \end{align}
    where $Q = Q^\mathrm{T} \succ 0$, $\tilde{u}_{k} = u_{k} - Uw_{k}$. $\hfill \square$
\end{problem}

Problem \ref{prb:1} is a standard discrete-time linear quadratic regulator problem. The solution to this problem is an optimal feedback controller of the form
\begin{equation}
    \tilde{u}_{k}^{\star} = -K^{\star}\tilde{\zeta}_{k}, \label{eq:optContErrSys}
\end{equation}
where $K^{\star} = (1+\bar{B}^\mathrm{T}P^{\star}\bar{B})^{-1}\bar{B}^\mathrm{T}P^{\star}\bar{A}$ and  $P^{\star}=P^{\star \mathrm{T}} \succ 0$ solves the following discrete-time algebraic Riccati equation
\begin{equation}
    \bar{A}^\mathrm{T}P^{\star}\bar{A}-P^{\star}+{Q}-\bar{A}^\mathrm{T}P^{\star}\bar{B}(1+\bar{B}^\mathrm{T}P^{\star}\bar{B})^{-1}\bar{B}^\mathrm{T}P^{\star}\bar{A} = 0. \label{eq:DTRic}
\end{equation}
The optimal controller for \eqref{eq:augSys1} can be obtained $\forall k \in \mathbb{Z}_{+}$ as 
\begin{equation}
    u_{k}^{\star} = \tilde{u}_{k}^{\star} + Uw_{k} \coloneqq -K_{x}^{\star}x_{k}-K_{z}^{\star}z_{k}. \label{eq:optCont}
\end{equation}

\section{Resilience analysis under DoS Attacks}\label{sec:resAnalysis}
In this paper, we examine scenarios where DoS attacks simultaneously impact both the measurement and control channels of the augmented system described by \eqref{eq:augSys1}. It is assumed that, during DoS attacks, the transmission and reception of data are both disrupted. Let $\{h_{m}\}_{m \in \mathbb{Z}_{+}}$ denote the sequence of off/on transitions of DoS, where $h_{0} \geq 0$. The $m^{\text{th}}$ DoS attack interval of length $\tau_{m}$ is represented as $\mathcal{J}_{m} \coloneqq [h_{m}, h_{m}+\tau_{m})$. For each interval $[k_{1},k_{2}]$, let $\Lambda_{N}(k_{1}, k_{2})$ and $\Lambda_{D}(k_{1}, k_{2})$ denote the set of time instants where communication is allowed and denied, respectively. Thus, $\Lambda_{N}(k_{1}, k_{2})$ and $\Lambda_{D}(k_{1}, k_{2})$ can be defined as follows
\begin{align}
    \Lambda_{D}(k_{1}, k_{2}) &\coloneqq \bigcup_{m \in \mathbb{Z}_{+}} \mathcal{J}_{m} \bigcap [k_{1}, k_{2}], \label{eq:gamD}\\
    \Lambda_{N}(k_{1}, k_{2}) &\coloneqq [k_{1}, k_{2}] \setminus \Lambda_{D}(k_{1}, k_{2}). \label{eq:gamN}
\end{align}
The following assumptions are made regarding DoS frequency and DoS duration.
\begin{assumption}(\textit{DoS Frequency})\label{assum:DoSFreq}
    There exist $\eta > 1$ and and $\tau_{D} > 0$ such that $\forall k_{2}>k_{1} \geq 0$,
    \begin{equation}
        n(k_{1}, k_{2}) \leq \eta + \frac{k_{2}-k_{1}}{\tau_{D}},\label{eq:DoSFreq}
    \end{equation}
    where $n(k_{1}, k_{2})$ denotes the number of DoS off/on transitions occurring on the interval $[k_{1},k_{2}]$. $\hfill \square$
\end{assumption}
\begin{assumption}(\textit{DoS Duration})\label{assum:DoSDur}
    There exist $T > 1$ and and $\kappa > 0$ such that $\forall k_{2}>k_{1} \geq 0$,
    \begin{equation}
        \lvert \Lambda_{D}(k_{1}, k_{2}) \rvert \leq \kappa + \frac{k_{2}-k_{1}}{T}, \label{eq:DoSDur}
    \end{equation}
    where $\lvert \Lambda_{D}(k_{1}, k_{2}) \rvert$ denotes the Lebesgue measure of the set $\Lambda_{D}(k_{1}, k_{2})$. $\hfill \square$
\end{assumption}

When the system is under DoS attack, the control input and internal model can be expressed $\forall k \in \mathbb{Z}_{+}$ as
\begin{align}
    u_{k} &= -K^{\star}\zeta_{k_{m(k)}}, \label{eq:optContDoS} \\
    z_{k+1} &= Ez_{k} + \mathcal{G}_{2}e_{k_{m(k)}}, \label{eq:intModDoS}
\end{align}
where $k_{m(k)}$ represents the most recent time instant at which the updated information is received. Let $\bar{\epsilon}_{k} = \zeta_{k_{m(k)}} - \zeta_{k}$ and $\underline{\epsilon}_{k} = e_{k_{m(k)}} - e_{k}$ be the error values between last successfully received values and actual values. Using the optimal controller \eqref{eq:optContDoS} and the internal model \eqref{eq:intModDoS}, the following closed-loop system is obtained
\begin{equation}
       \zeta_{k+1} = (\bar{A}-\bar{B}K^{\star})\zeta_{k} - \bar{B}K^{\star}\bar{\epsilon}_{k} + \bar{D}w_{k} +  \left[
        \begin{array}{c}
        0 \\
        \mathcal{G}_{2}\underline{\epsilon}_{k}\\
        \end{array}
        \right], \label{eq:augSysDoS}
\end{equation}
where $\zeta_{k} = [x_{k}^\mathrm{T}\quad z_{k}^\mathrm{T}]^\mathrm{T}$, $\bar{D} = [D^\mathrm{T}\quad(\mathcal{G}_{2}F)^{T}]^\mathrm{T}$ . Defining $\tilde{\zeta}_{k} = \zeta_{k} - \Xi w_{k}$, where $\Xi = [X^\mathrm{T}\quad Z^\mathrm{T}]^\mathrm{T}$, we have
\begin{equation}
    \bar{\epsilon}_{k} = \zeta_{k_{m(k)}} - \zeta_{k} = \tilde{\zeta}_{k_{m(k)}} - \tilde{\zeta}_{k} \label{eq:err1DoS}
\end{equation}
\begin{equation}
    \underline{\epsilon}_{k} = e_{k_{m(k)}} - e_{k} = \bar{C}\bar{\epsilon}_{k}. \label{eq:err2DoS}
\end{equation}
From \eqref{eq:augSysDoS}-\eqref{eq:err2DoS} the following error system can be obtained
\begin{align}
   \tilde{\zeta}_{k+1} &= (\bar{A}-\bar{B}K^{\star})\tilde{\zeta}_{k} - \bar{B}K^{\star}\bar{\epsilon}_{k} +  \left[
   \begin{array}{c}
   0 \\
   \mathcal{G}_{2}\bar{C}\bar{\epsilon}_{k}\\
   \end{array}
   \right],  \nonumber \\
   e_{k} &= \bar{C}\tilde{\zeta}_{k}. \label{eq:errSysDoS}
\end{align}
In this work, we seek to give an lower bound on the DoS duration parameter $T$, such that output regulation is achieved under DoS attacks. This is obtained in the following Theorem.
\begin{theorem}\label{th:4p1}
    The error system described in \eqref{eq:errSysDoS} is globally asymptotically stable if the following condition on the DoS duration parameter $T$ holds
    \begin{equation}
            T > 1 + \frac{\text{log}(1+\omega_{2})}{-\text{log}(1-\omega_{1})} \coloneqq T^{\star}, \label{eq:dosDurT}
    \end{equation}
    where 
    \begin{align}
        &\omega_{1} = \frac{\lambda_{m}(Q)}{\lambda_{M}(P^{\star})}, \omega_{2} = \frac{\alpha_{1}+4\alpha_{2}}{\lambda_{m}(P^{\star})}, \nonumber \\ 
        &\alpha_{1} = 1 + 2\lvert K^{\star\mathrm{T}}\bar{B}^\mathrm{T}P^{\star}\bar{B}K^{\star}\rvert^{2} + 2\lvert\bar{A}^\mathrm{T}P^{\star}\bar{A}\rvert^{2}, \nonumber \\ 
        &\alpha_{2} = 2 + 4 \lvert K^{\star\mathrm{T}}\bar{B}^\mathrm{T}P^{\star}\bar{B}K^{\star}\rvert^{2} + 4 \lvert\tilde{D}^\mathrm{T}P^{\star}\tilde{D}\rvert^{2}.\nonumber
    \end{align}
\end{theorem}
\begin{proof}
     Defining the Lyapunov function $V=\tilde{\zeta}_{k}^\mathrm{T}P^{\star}\tilde{\zeta}_{k}$, the following can be obtained
    \begin{align}
         V(\tilde{\zeta}_{k+1})-V(\tilde{\zeta}_{k}) &= \tilde{\zeta}_{k+1}^\mathrm{T}P^{\star}\tilde{\zeta}_{k+1}-\tilde{\zeta}_{k}^\mathrm{T}P^{\star}\tilde{\zeta}_{k} \nonumber\\
       \leq & -\lambda_{m}(Q)\lvert \tilde{\zeta}_{k} \rvert^{2} - \lvert \sqrt{P^\star}\bar{B}K^{\star}\tilde{\zeta}_{k} \nonumber \\
        -&\sqrt{P^\star}\bar{B}K^{\star}\bar{\epsilon}_{k}\rvert^{2} - \lvert \sqrt{P^\star}\bar{A}\tilde{\zeta}_{k}-\sqrt{P^\star}\tilde{D}\bar{\epsilon}_{k}\rvert^{2}\nonumber \\
       -&\lvert \sqrt{P^\star}\bar{B}K^{\star}\bar{\epsilon}_{k} + \sqrt{P^\star}\bar{A}\tilde{\zeta}_{k} \rvert^{2} \nonumber \\
       -& \lvert \sqrt{P^\star}\bar{B}K^{\star}\tilde{\zeta}_{k}+\sqrt{P^\star}\tilde{D}\bar{\epsilon}_{k} \rvert^{2}  \nonumber \\
       -& \lvert \sqrt{P^\star}\bar{B}K^\star \bar{\epsilon}_{k}+\sqrt{P^\star}\tilde{D} \bar{\epsilon}_{k}\rvert^{2} \nonumber \\
       +& 2\tilde{\zeta}_{k}^\mathrm{T}\bar{A}^\mathrm{T}P^\star\bar{A}\tilde{\zeta}_{k} + 2\tilde{\zeta}_{k}^\mathrm{T}K^{\star\mathrm{T}}\bar{B}^\mathrm{T}P^\star\bar{B}K^\star\tilde{\zeta}_{k} \nonumber \\
        +&4\bar{\epsilon}_{k}^\mathrm{T}K^{\star\mathrm{T}}\bar{B}^\mathrm{T}P^\star\bar{B}K^\star\bar{\epsilon}_{k} + 4\bar{\epsilon}_{k}^\mathrm{T}\tilde{D}^\mathrm{T}P^\star\tilde{D}\bar{\epsilon}_{k}, \label{eq:lyapIneqLm4p1}
    \end{align}
    where the inequality is obtained using completion of squares, $\tilde{D} = [0,~(\mathcal{G}_{2}\bar{C})^{\mathrm{T}}]^{\mathrm{T}}$. From \eqref{eq:lyapIneqLm4p1} and considering the interval $W_{m} \coloneqq [h_{m}+\tau_{m}, h_{m+1})$ where communications are normal, i.e., $\bar{\epsilon}_{k}=0$, we have $V(\tilde{\zeta}_{k+1})-V(\tilde{\zeta}_{k}) \leq -\lambda_{m}(Q)\lvert\tilde{\zeta}_{k}\rvert^{2}$, which implies
    \begin{equation}
        V(\tilde{\zeta}_{k}) \leq (1-\omega_{1})^{k-h_{m}-\tau_{m}}V(\tilde{\zeta}_{h_{m}+\tau_{m}}), \forall k \in W_{m},\label{eq:lyapIneqNoDoSLm4p1}
    \end{equation}
    During the interval $Z_{m}\coloneqq[h_{m}, h_{m}+\tau_{m})$ where communications are denied, the error is bounded by $\lvert \bar{\epsilon}_{k} \rvert \leq \lvert\tilde{\zeta}_{h_{m}}\rvert + \lvert\tilde{\zeta}_{k}\rvert $ and \eqref{eq:lyapIneqLm4p1} is equivalent to $V(\tilde{\zeta}_{k+1})-V(\tilde{\zeta}_{k}) \leq (\alpha_{1}+\alpha_{2}) \lvert\tilde{\zeta}_{k}\rvert^{2} + \alpha_{2} \lvert \tilde{\zeta}_{h_{m}}\rvert^{2} + 2\alpha_{2} \lvert\tilde{\zeta}_{h_{m}}\rvert\lvert \tilde{\zeta}_{k}\rvert$, which implies
    \begin{align}
       V(\tilde{\zeta}_{k+1})-V(\tilde{\zeta}_{k}) \leq \omega_{2}\text{max}\{V(\tilde{\zeta}_{k}),V(\tilde{\zeta}_{h_{m}})\},
    \end{align}
    Then, $\forall k\in Z_{m}$ we have
    \begin{align}
       V(\tilde{\zeta}_{k}) \leq (1+\omega_{2})^{k-h_{m}} V(\tilde{\zeta}_{h_{m}}). \label{eq:lyapIneqDoSLm4p1}
    \end{align}
    
    \begin{lemma}
        For all $k \in \mathbb{Z}_{+}$, $V(\tilde{\zeta}_{k})$ satisfies
        \begin{align}
            V(\tilde{\zeta}_{k}) \leq (1-\omega_{1})^{\lvert \Lambda_{N}(0,k) \rvert} (1+\omega_{2})^{\lvert \Lambda_{D}(0,k) \rvert} V(\tilde{\zeta}_{0}). \label{eq:lyapIneqCombLm4p1}
        \end{align}
    \end{lemma}
    \begin{proof}
        We use induction to prove the claim. Consider the interval $W_{-1} = [0,h_{0}]$. \eqref{eq:lyapIneqCombLm4p1} holds trivially if $h_{0} = 0$. If $h_{0}>0$, over $W_{-1}$, $V(\tilde{\zeta}_{k})$ obeys \eqref{eq:lyapIneqNoDoSLm4p1}. Thus, \eqref{eq:lyapIneqCombLm4p1} follows by noting that, $\lvert\Lambda_{N}(0,k)\rvert = k$ and $\lvert\Lambda_{D}(0,k)\rvert = 0$, $\forall k \in W_{-1}$. Next, assume that \eqref{eq:lyapIneqCombLm4p1} holds for the interval $[0, h_{p}], p\in\mathbb{Z}_{+}$. Then we have,
        \begin{align}
            V(\tilde{\zeta}_{h_{p}}) \leq (1-\omega_{1})^{\lvert \Lambda_{N}(0,h_{p}) \rvert} (1+\omega_{2})^{\lvert \Lambda_{D}(0,h_{p}) \rvert} V(\tilde{\zeta}_{0}), \label{eq:lyapIneqCombHypLm4p1} 
        \end{align}
        Next, consider the interval $Z_{p}\coloneqq [h_{p},h_{p}+\tau_{p})$. Then, over $Z_{p}$, $V(\tilde{\zeta}_{k})$ obeys \eqref{eq:lyapIneqDoSLm4p1} as follows
        \begin{equation}
            V(\tilde{\zeta}_{k}) \leq (1+\omega_{2})^{k-h_{p}} V(\tilde{\zeta}_{h_{p}}).\label{eq:lyapIneqDoSZpLm4p1}
        \end{equation}
        By substituting \eqref{eq:lyapIneqCombHypLm4p1} in \eqref{eq:lyapIneqDoSZpLm4p1}, \eqref{eq:lyapIneqCombLm4p1} follows by noting that $\lvert\Lambda_{N}(0,k)\rvert = \lvert\Lambda_{N}(0,h_{p})\rvert$, $\lvert\Lambda_{D}(0,k)\rvert = k-h_{p}+\lvert\Lambda_{D}(0,h_{p})\rvert$, $\forall k \in Z_{p}$. Therefore, \eqref{eq:lyapIneqCombLm4p1} holds for all $k \in [0, h_{p}+\tau_{p}]$.
        
        Next, consider the interval $W_{p}\coloneqq [h_{p}+\tau_{p}, h_{p+1})$. Then, over $W_{p}$, $V(\tilde{\zeta}_{k})$ obeys \eqref{eq:lyapIneqNoDoSLm4p1} as follows $\forall k \in W_{m}$.
        \begin{equation}
        V(\tilde{\zeta}_{k}) \leq (1-\omega_{1})^{k-h_{p}-\tau_{p}}V(\tilde{\zeta}_{h_{p}+\tau_{p}}). 
        \end{equation}
        In particular, we have,
        \begin{align}
        V(\tilde{\zeta}_{k}) &\leq (1-\omega_{1})^{k-h_{p}-\tau_{p}}(1+\omega_{2})^{\tau_{p}}V(\tilde{\zeta}_{h_{p}}). \label{eq:lyapIneqNoDoSWpLm4p1}
        \end{align}
        Then, by substituting \eqref{eq:lyapIneqCombHypLm4p1} in \eqref{eq:lyapIneqNoDoSWpLm4p1}, \eqref{eq:lyapIneqCombLm4p1} follows by noting that $\lvert \Lambda_{N}(0,k)\rvert = k-h_{p}-\tau_{p}+\lvert\Lambda_{N}(0,h_{p})\rvert$, $\lvert \Lambda_{D}(0,k)\rvert = \tau_{p}+\lvert\Lambda_{D}(0,h_{p})\rvert$, $\forall k \in W_{p}$. Therefore, \eqref{eq:lyapIneqCombLm4p1} holds for all $k \in [0, h_{p+1}]$, where $p \in \mathbb{Z}_{+}$.
    \end{proof}
    Note that $\lvert\Lambda_{N}(0, k)\rvert = k-\lvert\Lambda_{D}(0,k)\rvert$.  Then, from \eqref{eq:lyapIneqCombLm4p1} and Assumption~\ref{assum:DoSDur}, we have
    \begin{align}
        (1-\omega_{1})^{\lvert \Lambda_{N}(0,k) \rvert} (1+\omega_{2})^{\lvert \Lambda_{D}(0,k) \rvert} \leq \bigg[\frac{1+\omega_{2}}{1-\omega_{1}}\bigg]^{\kappa} \Delta^{k},\label{eq:convProof1stTerm}
    \end{align}
    where $\Delta = (1-\omega_{1})^{\frac{T-1}{T}}(1+\omega_{2})^{\frac{1}{T}}$. Under the condition \eqref{eq:dosDurT}, $\Delta < 1$ (for example, choose $T = \frac{T^{\star}}{\delta}$, where $\delta \in (0,1)$). 
    
    Using \eqref{eq:convProof1stTerm}, the following can be obtained from \eqref{eq:lyapIneqCombLm4p1}
    \begin{align}
        V(\tilde{\zeta}_{k}) &\leq \bigg[\frac{1+\omega_{2}}{1-\omega_{1}}\bigg]^{\kappa} \Delta^{k} V(\tilde{\zeta}_{0}).
        \label{eq:lyapIneqCombSimLm4p1}
    \end{align}
    Thus, the following can be obtained from \eqref{eq:lyapIneqCombSimLm4p1} and \eqref{eq:errSysDoS}
    \begin{align}
        \lvert\tilde{\zeta}_{k}\rvert &\leq \beta_{\tilde{\zeta}}(\lvert \tilde{\zeta}_{0} \rvert, k),\label{eq:ISSCombLm4p1} \\
        \lvert e_{k}\rvert &\leq \beta_{e}(\lvert \tilde{\zeta}_{0} \rvert, k),\label{eq:IOSCombLm4p1}
    \end{align}
    where 
    
    $\beta_{\tilde{\zeta}}(\lvert \tilde{\zeta}_{0} \rvert, k) = \sqrt{\bigg[\frac{1+\omega_{2}}{1-\omega_{1}}\bigg]^{\kappa}\frac{\lambda_{M}(P^{\star})}{\lambda_{m}(P^{\star})}\Delta^{k}}  \lvert\tilde{\zeta}_{0}\rvert$ and 
    
    $\beta_{e}(\lvert \tilde{\zeta}_{0} \rvert, k) = \lvert\bar{C}\rvert\sqrt{\bigg[\frac{1+\omega_{2}}{1-\omega_{1}}\bigg]^{\kappa}\frac{\lambda_{M}(P^{\star})}{\lambda_{m}(P^{\star})}\Delta^{k}} \lvert\tilde{\zeta}_{0}\rvert$, 
    
    are class $\mathcal{KL}$ functions. From \eqref{eq:ISSCombLm4p1} it is clear that \eqref{eq:errSysDoS} has global asymptotic stability property. Thus, we have $\lim_{k\rightarrow\infty}(x_{k}-Xw_{k}) = 0$, and $\lim_{k\rightarrow\infty}e_{k} = 0$. This implies asymptotic tracking and disturbance rejection. The proof is thus complete.
\end{proof}

\section{Learning-based design under DoS Attacks}\label{sec:learn}
In this section, we propose an online strategy to learn the optimal controller \eqref{eq:optCont} while the system is under DoS attacks. We assume that the system matrices $A, B$ and $D$ are unknown. We use policy iteration to learn the optimal controller. The idea of policy iteration is to implement both policy evaluation \cite{hewer1971iterative}
\begin{equation}
    \bar{A}_{j}^\mathrm{T}P_{j}\bar{A}_{j}-P_{j}+{Q}+K_{j}^{\mathrm{T}}K_{j} = 0 \label{eq:LyapPI}
\end{equation}
and policy improvement
\begin{equation}
    K_{j+1} = (1 + \bar{B}^\mathrm{T}{P}_{j}\bar{B})^{-1}\bar{B}^\mathrm{T}{P}_{j}\bar{A},
\end{equation}
where $\bar{A}_{j} = \bar{A}-\bar{B}K_{j}$

Firstly, we rewrite the augmented system \eqref{eq:augSys1} as
\begin{equation}
    \zeta_{k+1} =\bar{A}_{j}\zeta_{k} + \bar{B}(u_{k}+K_{j}\zeta_{k}) + \bar{D}w_{k}, \label{eq:augSysLearn}
\end{equation}
Along the trajectories of (\ref{eq:augSysLearn}), one can obtain that:
\begin{align}
&\zeta_{k+1}^\mathrm{T}P_{j}\zeta_{k+1} -\zeta_{k}^\mathrm{T}P_{j}\zeta_{k}=\big[\bar{A}_{j}\zeta_{k} + \bar{B}(u_{k}+K_{j}\zeta_{k}) \nonumber \\
&+ \bar{D}w_{k}\big]^\mathrm{T}P_{j}\big[\bar{A}_{j}\zeta_{k} + \bar{B}(u_{k}+K_{j}\zeta_{k}) + \bar{D}w_{k}\big]- \zeta_{k}^\mathrm{T}P_{j}\zeta_{k}.
\end{align}
Then, using \eqref{eq:LyapPI}, we have:
\begin{align}
&\zeta_{k+1}^\mathrm{T}P_{j}\zeta_{k+1} - \zeta_{k}^\mathrm{T}P_{j}\zeta_{k} + \zeta_{k}^\mathrm{T}Q_{j}\zeta_{k} = 2 \zeta_{k}^\mathrm{T}\Gamma_{1j}^\mathrm{T}u_{k} \nonumber \\
&+ 2\zeta_{k}^\mathrm{T}\Gamma_{1j}^\mathrm{T}K_{j}\zeta_{k}  -\zeta_{k}^\mathrm{T}K_{j}^\mathrm{T}\Gamma_{2j}K_{j}\zeta_{k} + u_{k}^\mathrm{T}\Gamma_{2j}u_{k} \nonumber \\
&+2\zeta_{k}^\mathrm{T}\Theta_{1j}w_{k} + 2 u_{k}^\mathrm{T}\Theta_{2j}w_{k}+  w_{k}^\mathrm{T}\Theta_{3j}w_{k}, \label{eq:dataDriven}
\end{align}
where
$Q_{j} = Q+K_{j}^\mathrm{T}K_{j}$, $\Theta_{1j} = \bar{A}^\mathrm{T}P_j\bar{D}$, $\Theta_{2j} = \bar{B}^\mathrm{T}P_{j}\bar{D}$, $\Theta_{3j} = \bar{D}^\mathrm{T}P_{j}\bar{D}$, $\Gamma_{1j} = \bar{B}^\mathrm{T}P_{j}\bar{A}$, $\Gamma_{2j} = \bar{B}^\mathrm{T}P_{j}\bar{B}$. By Assumption~\ref{assum:DoSDur}, there always exists a sequence $\{k_{s}\}_{s=0}^{\infty}$ such that communications are allowed. Then, by collecting online data, the following linear equation can be obtained from \eqref{eq:dataDriven}
\begin{equation}
   \Psi_{j}\theta_{j} = -J_{\zeta,\zeta}\text{vec}(Q_{j}),\label{eq:LeastSquaresPI}
\end{equation} 
where 

$\Psi_{j}=\big[\Xi_{\zeta},-2J_{\zeta,u}-2J_{\zeta, \zeta}(I_{n} \otimes K_{j}^\mathrm{T}), J_{K_{j}\zeta} - J_{u}, -2J_{w,\zeta},\\-2J_{w,u},-J_{w}\big]$, 
$\theta_{j} = \big[\text{vecs}(P_{j})^\mathrm{T}, \text{vec}(\Gamma_{1j})^\mathrm{T}, \text{vecs}(\Gamma_{2j})^\mathrm{T}, \\ \text{vec}(\Theta_{1j})^\mathrm{T}, \text{vec}(\Theta_{2j})^\mathrm{T}, \text{vecs}(\Theta_{3j})^\mathrm{T}\big]^\mathrm{T}$. 

One can solve \eqref{eq:LeastSquaresPI} in the least square sense. Under certain choices of $E$, the matrix $J_{w}$ may not be full column rank. In such cases, it becomes necessary to reduce the columns of $J_{w}$ such that $\Psi_{j}$ is full column rank (see \cite{chakraborty2023adaptive}, \cite{chakraborty2023learning}). Denote $\bar{\Psi}_{j}$ as the matrix which contains the reduced columns of $J_{w}$ such that $\bar{\Psi}_{j}$ is full column rank. Since $\bar{\Psi}_{j}$ has less number of columns, the size of $\theta_{j}$ is also reduced, which is denoted as $\bar{\theta}_{j}$. Then, the least squares problem \eqref{eq:LeastSquaresPI} can be written as 
\begin{equation}
   \bar{\Psi}_{j}\bar{\theta}_{j} = -J_{\zeta, \zeta}\text{vec}(Q_{j}).\label{eq:LeastSquaresPI1}
\end{equation} 

\begin{assumption}\label{rankCondPhase1}
There exists a $s^\star \in \mathbb{Z}_{+}$ such that for all $s>s^\star$, and for any sequence $k_{0} < k_{1} < \cdots < k_{s}$:
\begin{align}\label{rankCond}
    &\mathrm{rank}([J_{\zeta}, J_{\zeta,u}, J_{u}, J_{w,\zeta}, J_{w,u}, J_{w}]) = \frac{n(n+1)}{2} + n \nonumber \\
    &+ 1 + nq + q + \frac{q(q+1)}{2}-N,
\end{align}
where $N$ is the number of linearly dependent columns of $J_{w}$. $\hfill \square$
\end{assumption}
\begin{remark}
A typical choice of $s^\star$ can be $s^\star\geq\frac{n(n+1)}{2} + n + 1 + nq + q + \frac{q(q+1)}{2}$. \hfill  $\square$
\end{remark}
\begin{remark}
    Under Assumption~\ref{rankCondPhase1}, \eqref{eq:LeastSquaresPI1} has a unique solution and the sequences $\{P_{j}\}_{j=0}^{\infty}$ and $\{K_{j}\}_{j=0}^{\infty}$ obtained using Algorithm~\ref{algo: phase 1 ADP} converge to a neighborhood of the optimal values $P^\star$ and $K^\star$, respectively \cite{chakraborty2023adaptive}, \cite{chakraborty2023learning}. \hfill  $\square$
\end{remark}
\begin{algorithm}
	\caption{Online Model-free Policy Iteration}\label{algo: phase 1 ADP}
	\begin{algorithmic}[1]
	\State Employ $u_{k} = -K_{0}\zeta_{k} + \eta_{k}$ as the input on the time horizon [$k_{0}$, $k_{s}$], where $K_{0}$ is initial stabilizing gain and $\eta_{k}$ is the exploration signal. 
	\State Compute $\bar{\Psi}_{j}$ until the rank condition in (\ref{rankCond}) is satisfied. Let $j=0$.
	\State Solve for $\bar{\theta}_{j}$ from \eqref{eq:LeastSquaresPI1}.
    \State Compute $K_{j+1} = (1 + \Gamma_{2j})^{-1}\Gamma_{1j}$.
	\State Let $j \leftarrow j+1$ and repeat Step 3. until $\lVert P_{j} - P_{j-1} \rVert \leq \epsilon_{0}$ for $j \geq 1$, where $\epsilon_{0} > 0$ is a predefined small threshold. 
	\end{algorithmic}
\end{algorithm}
\begin{figure}[tbh!]
    \centering
  {\includegraphics[width=0.4\textwidth,trim={0 0 0 0},clip]{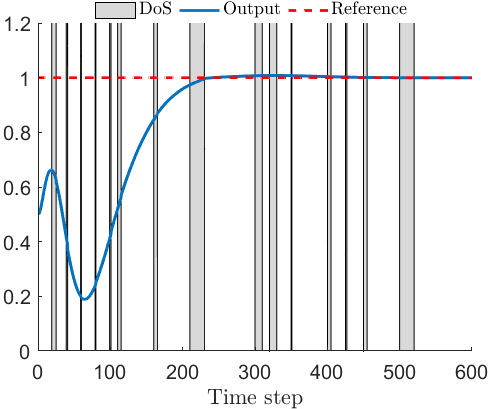}}
  \caption{Tracking and disturbance rejection under DoS attacks.}
  \label{Output}
\end{figure}
\begin{figure}[tbh!]
    \centering
  {\includegraphics[width=0.4\textwidth,trim={0 0 0 0},clip]{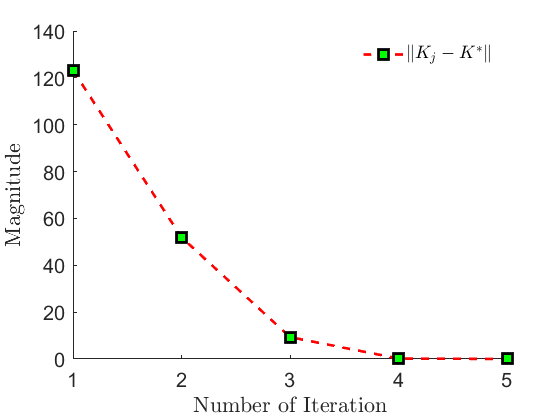}}
  \caption{Convergence of $K_{j}$ to $K^\star$.}
  \label{KjKStar}
\end{figure}
\begin{figure}[tbh!]
    \centering
  {\includegraphics[width=0.4\textwidth,trim={0 0 0 0},clip]{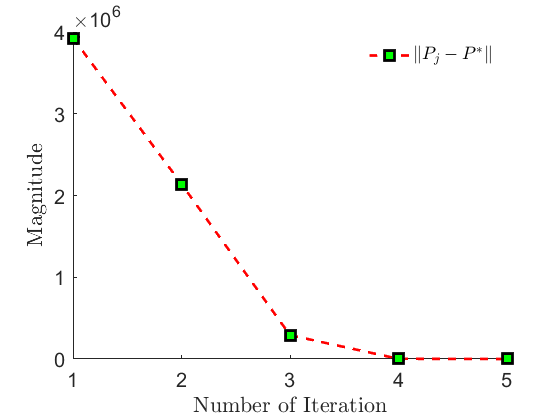}}
  \caption{Convergence of $P_{j}$ to $P^\star$.}
  \label{PjPStar}
\end{figure}

\section{Simulation Results and Discussion}\label{sec:sim}
In this section, we show the efficacy of the proposed algorithm by applying it to an inverted pendulum on a cart with the following system matrices,

$ A = \begin{bmatrix}
    1 & T & 0 & 0\\
    0 & 1-\frac{bT}{M} & -\frac{mgT}{M} & 0\\
    0 & 0 & 1 & T\\
    0 & \frac{bT}{lM} & \frac{(M+m)gT}{lM} & 1
    \end{bmatrix}$,
$B = \begin{bmatrix}
    0\\
    \frac{T}{M}\\
    0\\
    -\frac{T}{lM}
\end{bmatrix}$,\\

$D = \begin{bmatrix}
    0 & 0.01 & 0 & 0.01
    \end{bmatrix}^\mathrm{T}$,
$E = 1$, $C = [1~0~0~0]$, $F = -1$, $\mathcal{G}_{2} = 0.5$.
For the meaning and value of the parameters please refer to \cite{gurumoorthy1993controlling}. The initial conditions are given as $x_{0} = [0.5,~0,~0,~0]$, and $w_{0} = 1$. The weight matrices are chosen as $Q = \text{diag}(1000, 1000, 1000, 1000, 15)$, and $R = 1$. The DoS parameters are selected as $\kappa = 40$, $\tau_{D} = 15$, $T = 10$, and $\eta = 1$. The exploration signal in Algorithm~\ref{algo: phase 1 ADP} is chosen as the summation of sinusoidal waves with different frequencies. Using input-state data for $k\in[0,100]$, Algorithm~\ref{algo: phase 1 ADP} converges with a tolerance of $\epsilon_{0}=0.5$ to a neighborhood of the optimal values $P^\star$ and $K^\star$ in five iterations as shown in Figs.~\ref{KjKStar}, and~\ref{PjPStar}. The optimal controller gain $K^\star$ and the controller gain learned using Algorithm~\ref{algo: phase 1 ADP} are given in Table~\ref{tab:compKKStar}.
\begin{table}[ht]
\centering
\caption{Comparison of controller gain values.}
\label{tab:compKKStar}
\begin{tabular}{@{}c|cccccc@{}}
\toprule
 & & & Index & &   \\ 
\toprule
Controller & 1 & 2 & 3 & 4 & 5\\ 
\midrule
$K^\star$ &-153.9801 &-99.7489 &-283.9957 &-56.1038 &-2.6548 \\
\midrule
$K_{8}$ &-153.9802 &-99.7490 &-283.9958 &-56.1038 &-2.6548\\ 
\bottomrule
\end{tabular}
\end{table}


We immediately apply the learned controller after $k = 100$. Fig.~\ref{Output} shows the output and reference trajectories, with the DoS attacks represented as shaded areas. The learned controller can track the reference signal even in the presence of DoS attacks. The DoS duration parameter can be obtained as $T^{\star} = 6.3487\times 10^{6}$. Similar to \cite{de2015input}, \cite{gao2022resilient}, these are sufficient conditions to guarantee the resilience and stability of the closed-loop system. In practice $T^{\star}$ can be much smaller. This is demonstrated by applying stronger DoS attacks after 100 time steps.
\section{Conclusion and Future Works}\label{sec:conc}
This paper investigates the challenge of achieving optimal output regulation of discrete-time linear systems with unknown parameters while facing denial-of-service (DoS) attacks. We have proposed a resilient online policy iteration algorithm capable of learning the optimal controller using only the input-state data in the presence of DoS attacks. An upper bound on the DoS duration is achieved to guarantee the stability of the closed-loop system. Finally, the proposed technique is applied to an inverted pendulum on a cart. 

Future work will focus on extending this technique to discrete-time non-linear systems.

\balance
\bibliographystyle{IEEEtran}        
\bibliography{IEEEabrv, ref} 





\end{document}